\documentclass[12pt]{article}
\usepackage{euscript,amsmath,amssymb,latexsym}

\newcommand{\be}{\begin{equation}}
\newcommand{\ee}{\end{equation}}
\newcommand{\bea}{\begin{eqnarray}}
\newcommand{\eea}{\end{eqnarray}}

\begin{document}
\title{
\vspace{1cm} {\bf Time Irreversibility Problem and \\ Functional
Formulation of Classical Mechanics}
 }
\author{Igor V. Volovich
 \\
{\it  Steklov Mathematical Institute}
\\ {\it Gubkin St.8, 119991 Moscow, Russia}
\\ email:\,volovich@mi.ras.ru}

\date {~}
\maketitle

\begin{abstract}
The time irreversibility problem is the dichotomy of the reversible
microscopic dynamics and the irreversible macroscopic physics. This
problem was considered by Boltzmann, Poincar\'e, Bogolyubov and many
other authors and though some researchers claim that the problem is
solved, it deserves a further study. In this paper an attempt is
performed of the following solution of the irreversibility problem:
a formulation of microscopic dynamics is suggested which is
irreversible in time. In this way the contradiction between the
reversibility of microscopic dynamics and irreversibility of
macroscopic dynamics is avoided since both dynamics in the proposed
approach are irreversible.

A widely used notion of microscopic state of the system at a given
moment of time as a point in the phase space and also a notion of
trajectory and microscopic equation of motion does not have an
immediate physical meaning since arbitrary real numbers are non
observable. In the approach presented in this paper the physical
meaning is attributed not to an individual trajectory but only to a
bunch of trajectories or to the distribution function on the phase
space.

The fundamental equation of the microscopic dynamics in the proposed
``functional" approach is not the Newton equation but the Liouville
equation for the distribution function of the single particle.
Solutions of the Liouville equation have the property of
delocalization which accounts for irreversibility. It is shown that
the Newton equation in this approach appears as an approximate
equation describing the dynamics of the average values of the
position and momenta for not too long time intervals. Corrections to
the Newton equation are computed.
\end{abstract}
\maketitle

\newpage

\section{Introduction}
The time irreversibility problem is the problem of how to explain
the irreversible behaviour of macroscopic systems from the
time-symmetric microscopic laws.  The problem has been discussed by
 Boltzmann, Poincar\'e, Bogolyubov, Kolmogorov, von Neumann,
Landau, Prigogine, Feynman and many other authors \cite{Bol1} -
\cite{Bri} and it deserves a further study.

In particular, in works by Poincar\'e \cite{Poi1}, Landau and
Lifshiz \cite{LL}, Prigogine \cite{Pri},  Ginzburg \cite{Gin},
Feynman \cite{Fey} it is stressed that the irreversibility problem
is still an open problem. Poincar\'e \cite{Poi1} said that perhaps
we will never solve the irreversibility problem. Landau and Lifshiz
write about the principle of increasing entropy
  \cite{LL}: ``Currently it is not clear whether the
  law of increasing entropy  can be in principle derived from classical
  mechanics." Landau speculated that to explain
  the second law of thermodynamics one has to use quantum mechanical
  measurement arguments.

 From the other side Lebowitz \cite{Leb}, Goldstein \cite{Gol}
 and Bricmont \cite{Bri} state
that the irreversibility problem was basically solved already by
Boltzmann by using his notion of  macroscopic entropy and the
probabilistic approach.

 The microscopic mechanical description
of a system assumes that the state of the system at a given moment
of time is represented by a point in the phase space with an
invariant measure and the dynamics of the system is described by a
trajectory in the phase space, see \cite{LL, Kol, Arn, DNF, Ano,
Sin}. It is assumed that the microscopic laws of motion are known
(Newton or Schrodinger equations) and there is a problem of
derivation from them the macroscopic (Boltzmann, Navier-Stokes,...)
equations, see for example, \cite{LL, Zub}.

 There are  well known
critical remarks by Loschmidt and Poincar\'e and Zermelo on the
Boltzmann approach to the irreversibility problem and the
$H$-theorem. Loschmidt remarked that from the symmetry of the Newton
equations upon the reverse of time it follows that to every motion
of the system on the trajectory towards the equilibrium state one
can put into correspondence the motion out of the equilibrium state
if we reverse the velocities at some time moment. Such a motion is
in contradiction with the tendency of the system to go to the
equilibrium state and with the law of increasing of entropy.

Then, there is the Poincar\'e recurrence theorem which says that a
trajectory of a bounded isolated mechanical system will be many
times come to a  very  small neighborhood of an initial point. This
is also in contradiction with the motion to the equilibrium state.
This is the Poincar\'e--Zermelo paradox.

Boltzmann \cite{Bol5} gave the following answer to the Loschmidt
argument: ``We do not have to assume a special type of initial
condition in order to give a mechanical proof of the second law, if
we are willing to accept a statistical viewpoint. While any
individual non-uniform state (corresponding to low entropy) has the
same probability as any individual uniform state (corresponding to
high entropy), there are many more uniform states than non-uniform
states. Consequently, if the initial state is chosen at random, the
system is almost certain to evolve into a uniform state, and entropy
is almost certain to increase."

So, the answer by Boltzmann to the objection of Loschmidt was that,
firstly, the probabilistic considerations has been involved, and
secondly, he argued that with the overwhelming probability  the
evolution of the system will be occur in the direction of flow of
time, corresponding to the increasing entropy, since  there are many
more uniform states than non-uniform states. The answer by Boltzmann
to the Poincar\'e -- Zermelo objection was in the pointing out the
extremely long   Poincar\'e recurrence time.

These Boltzmann`s responses  are not very  convincing, from our
point of view, despite their vigorous support in recent works \cite
{Leb, Gol, Bri}. Involvement of probability considerations alone
does not clarify the issue of irreversibility, because if there is
symmetry in relation to the direction of time, it remains unclear
why the evolution in one direction is more likely than the other.

Then, the argument, that  there are many more uniform states than
non-uniform states does not clarify the issue of the dynamical
evolution since the dynamics does depend  on the form of the
potential energy between particles and for many potentials the
argument is simply wrong. Therefore this general   Boltzmann's
argument does not give a real insight to the irreversibility
problem.

Actually, Boltzmann in \cite{Bol5} considered ``a large but not
infinite number of absolutely elastic spheres, which move in a
closed container whose walls are completely rigid and likewise
absolutely elastic. No external forces act on our spheres.'' Even
for this simple model it is very difficult to make the Boltzmann
argument convincing, i.e. to get a mathematical result, see
\cite{Arn,Sin}.

Further, an indication to the extremely long Poincar\'e recurrence
time does not remove the contradiction between microscopic
reversibility and macroscopic irreversibility, and moreover no clear
mechanism for relaxation to equilibrium is presented.

Lebowitz  advanced \cite{Leb}, following to Boltzmann,
 the following arguments  to explain
irreversibility: a) the great disparity between microscopic and
macroscopic scales, b) a low entropy state of the early universe,
and c) the fact that what we observe is the behaviour of systems
coming from such an initial state -- not all possible systems.

From our viewpoint these arguments do not lead to explanation of
irreversibility even though it is said in \cite{Leb} that ``common
alternative explanations, such as those based on the ergodic or
mixing properties of probability distribution\,...\,are either
unnecessary, misguided or misleading'' .

Boltzmann proposed that we and our observed low-entropy world are a
random fluctuation in a higher-entropy universe. These cosmological
considerations of the early universe might be entertaining but they
should be related with the modern Friedmann \cite{Fri,Lin}
gravitational picture of the Big Bang and, what is most important,
there is no evidence that the irreversible behaviour of gas in a box
is related somehow with conditions in the early universe 14 billions
years ago.

Notice that in \cite{NV} it is shown that the Hawking black hole
information paradox is a special case of the irreversibility
problem.

Goldstein  said in \cite{Gol}:``The most famous criticisms of
Boltzmann's later works on the subject have little merit. Most
twentieth century innovations -- such as the identification of the
state of a physical system with a probability distribution $\rho$ on
its phase space, of its thermodynamic entropy with the Gibbs entropy
of $\rho$, and the invocation of the notions of ergodicity and
mixing for the justification of the foundations of statistical
mechanics -- are thoroughly misguided."

And then: ``This use of ergodicity is thoroughly misguided.
Boltzmann's key insight was that, given the energy of a system, the
overwhelming majority of its phase points on the corresponding
energy surface are equilibrium points, all of which look
macroscopically more or less the same.''

The Boltzmann argument about  ``the overwhelming majority'' (i.e.
``many more uniform states'') was discussed above.  Moreover, the
main point of the current  paper is that we shall use the
probability distribution and the Liouville equation not only in
statistical mechanics but also in classical mechanics, even for a
single particle in empty space.

A powerful method for obtaining kinetic equations from the Newton --
Liouville equations
 was developed  by Bogolyubov \cite {Bog}. He has considered infinite
 number of particles in infinite volume and
postulated the  condition of weakening of initial correlations
between particles in the distant past, through which the
irreversibility entered into the equation for the distribution
functions, as well as using a formal expansion in powers of density,
which leads to divergences.

Poincar\'e considered the model of free motion of gas particles in a
box with reflecting walls and showed that for solutions of the
Liouville equation in this model  there is,  in some sense, an
irreversible diffusion \cite {Poi3}. This result of Poincar\'e was
introduced to modern scientific literature by Kozlov, see \cite
{Koz1}, where the result of Poincar\'e was significantly
strengthened and consolidated. In the works of Kozlov  a method of
the weak limit in the nonequilibrium statistical mechanics has been
developed, and, in particular, it was proved that for some models
the system in the sense of weak convergence  tends to one and the
same limit in the past and in the future \cite {Koz1, Koz2}. The
method of the weak limit of \cite {Koz1, Koz2} had a significant
influence to the formulation of  the approach to the problem of
irreversibility through functional formulation of classical
mechanics.

Note that the stochastic limit \cite{ALV} gives a systematic method
for investigation of irreversible processes.

Questions about the increase of the fine and coarse entropies are
discussed in \cite {Poi2, LL, Che, Kac, Zub, KT, Koz2}.

In this paper we attempt to suggest the following approach to the
irreversibility problem and to paradoxes of Loschmidt and Poincar\'e
-- Zermelo: we propose a {\it formulation of  microscopic dynamics
which is irreversible in time}. Thus  the contradiction between
microscopic reversibility and macroscopic irreversibility of the
dynamics disappears, since both microscopic and macroscopic dynamics
in the proposed approach are irreversible.

Note that the conventional widely used concept of the microscopic
state of the system at some  moment in time  as the point in phase
space, as well as the notion of trajectory and the microscopic
equations of motion have no direct physical meaning, since arbitrary
real numbers not observable (observable physical quantities are only
presented by rational numbers, cf. the discussion of concepts of
space and time in \cite{ Vol1, Vol2, Man, VVZ, Khr1, DKKV, Var,
Zel}).

In the proposed ``functional" approach, the physical meaning is
attached not to a single trajectory, but only to a ``beam" of
trajectories, or the distribution function on phase space.
Individual trajectories are not observable, they could be considered
as ``hidden variables", if one uses the quantum mechanical notions,
see
 \cite{Vol3, Vol5}.

The fundamental equation of the microscopic dynamics of the proposed
functional probabilistic approach is not Newton's equation, but a
Liouville equation for distribution function. It is well known that
the Liouville equation is used in statistical mechanics for the
description of the motions of gas. Let us stress that we shall use
the Liouville equation for the description of a single particle in
the empty space.

Although the Liouville equation is symmetric in relation to the
reversion of  time, but his solutions have the property of {\it
delocalization}, that, generally speaking, can be interpreted as a
manifestation of irreversibility. It is understood that if at some
moment in time the distribution function describes a particle,
localized to a certain extent,  then over time the degree of
localization decreases, there is the spreading of distribution
function. Delocalization takes place even for a free particle in
infinite space, where there is no ergodicity and mixing.

In the functional approach to classical mechanics we do not derive
the statistical or chaotic properties of deterministic dynamics, but
we suggest that the Laplace's determinism  at the fundamental level
is absent not only in quantum, but also in classical mechanics.

We show that Newton's equation in the proposed approach appears as
an approximate equation describing the dynamics of the average
values of coordinates and momenta for not too long  time. We
calculate corrections to Newton's equation.

In the next section the fundamentals of the functional formulation
of classical mechanics are presented. Sections 3 and 4 deal with the
free movement of particles and Newton's equation for the average
coordinates. Comparison with quantum mechanics is discussed in
Section 5. General comments on the Liouville and Newton equations
are contained in section 6. Corrections to the Newton equation for a
nonlinear system are calculated in Section 7. Reversibility of
motion in classical mechanics and irreversibility in the functional
approach to the mechanics discussed in section 8. The dynamics of
the classical and quantum particle in a box and their
interrelationships are summarized in section 9.

\section{States and Observables in  \\ Functional Classical Mechanics}

Usually in classical mechanics the motion of a point body is
described by the  trajectory in the phase space, i.e. the values of
the coordinates and momenta as functions of time, which are
solutions of the equations of Newton or Hamilton.

Note, however, that this mathematical model is an idealization of
the physical process, rather far separated from reality. The
physical body always has the spatial dimensions, so a mathematical
point gives only an approximate description of the physical body.
The mathematical notion of a trajectory does not have direct
physical meaning, since it uses  arbitrary real numbers, i.e.
infinite decimal expansions, while the observation is only possible,
in the best case, of rational numbers, and even them only with some
error. Therefore, in the proposed ``functional" approach to
classical mechanics, we are not starting from Newton's equation, but
with the Liouville equation.

Consider the motion of a classical particle along a straight line in
the potential field. The general case of many particles in the
3-dimensional space is discussed below. Let $(q, p)$ be co-ordinates
on the plane $\mathbb{R}^2$  (phase space), $t\in\mathbb{R}$ is
time. The state of a classical particle at time $t$ will be
described by the function $\rho=\rho (q, p, t)$, it is the density
of the probability that the particle at time $t$ has the coordinate
$q$ and momentum $p$.

Note that the description of a mechanical system with the help of
probability distribution function $\rho=\rho (q, p, t)$ does not
necessarily mean that we are dealing with a set of identically
prepared ensemble of particles. Usually in probability theory one
considers  an ensemble of events  and a sample space
\cite{PR,BS,Khr2}. But we can use the description with the function
$\rho=\rho (q, p, t)$ also for individual bodies, such as planets in
astronomy (the phase space in this case the 6-dimensional). In this
case one can think on the ``ensemble" of different astronomers which
observe the planet, or on the  ``ensemble" of different models of
behaviour of a given object. Actually, it is implicitly always dealt
with the function $\rho=\rho (q, p, t)$ which takes into account the
inherent uncertainty in the coordinates and momentum of the body. An
application of these remarks to quantum mechanics will be discussed
in a separate work.

The specific type of function $\rho$ depends on the method of
preparation of the state of a classical particle at the initial time
and the type of potential field. When $\rho=\rho (q, p, t)$ has
sharp peaks at $q = q_0$ and $p = p_0$, we say that the particle has
the approximate values of coordinate and momentum $q_0$ and $p_0$.

Emphasize that  the exact derivation of the coordinate and momentum
can not be done, not only in quantum mechanics, where there is the
Heisenberg uncertainty relation, but also in classical mechanics.
Always there are some errors in setting the coordinates and momenta.
The concept of arbitrary real numbers, given by the infinite decimal
series, is a mathematical idealization, such numbers can not be
measured in the experiment.

Therefore, in the functional approach to classical mechanics the
concept of precise trajectory of a particle is absent, the
fundamental concept is a distribution function $\rho=\rho (q, p, t)$
and $\delta$-function as a distribution function is not allowed.

We assume that the continuously differentiable and integrable
function $\rho=\rho (q, p, t)$ satisfies the conditions:
\begin{equation}\label{rho}
\rho \geq 0,~~\int_{\mathbb{R}^2}\rho (q,p,t)dqdp=1,~ t\in
\mathbb{R}\,.
\end{equation}
The formulation of classical mechanics in the language of states and
observables is considered in \cite{ Mac, FY, Kli}. The functional
approach to classical mechanics differs  in the following respects.
Because the exact trajectory of a particle in the functional
approach does not exist, then the function $\rho=\rho (q, p, t)$ can
not be an arbitrary generalized function, it is the usual function
of class $L^1(\mathbb{R}^2)$, or even continuously differentiable
and integrable function.

In addition, the motion of particles in the functional approach is
not described directly by the Newton (Hamilton) equation. Newton's
equation in the functional approach is an approximate equation for
the average coordinates of the particles, and for non-linear
dynamics  there are  corrections to the Newton equations.

As is known, the mathematical description of a moving fluid or gas
is given by means of the density distribution functions $\rho
(q,t)$, as well as the velocity $v (q, t)$ and pressure $p (q, t)$,
see, for example, \cite{ LL2}. Let the function $\rho (q,p,t)$
describes a particle, as proposed in the functional formulation of
classical mechanics, and we set $\rho_c (q,t)=\int \rho (q,p,t)dp$.
We could ask the question can we determine by the form  of functions
$\rho (q,t)$ and $\rho_c (q,t)$ whether we are dealing with a
continuous medium or with a particle? The general answer is the
following: functions $\rho (q,t)$ and $\rho_c (q,t)$ satisfy
different equations (the Navier-Stokes  or Liouville equation) and
different conditions of normalization.

Note, however, that if an error in determining the coordinates and
momentum of particles is large enough, it really is not so easy to
determine, we have a case of, say, a fast-moving particle in a box
with reflecting walls, either a gas of  particles.

If $f = f (q, p)$  is a function on phase space, the average value
of $f$ at time $t$ is given by the integral
\begin{equation}\label{int1}
\overline{f}(t)=\int f(q,p)\rho (q,p,t)dqdp\,.
\end{equation}
In a sense we are dealing with a random process $\xi(t)$ with values
in the phase space. Motion of a point  body along a straight line in
the potential field will be described by the equation
\begin{equation}\label{Lio}
\frac{\partial \rho}{\partial t}=-\frac{p}{m}\frac{\partial
\rho}{\partial q}+\frac{\partial V(q)}{\partial q} \frac{\partial
\rho}{\partial p}\,.
\end{equation}
Here $V(q)$  is the potential field and mass $m>0$.

Equation (\ref{Lio}) looks like the Liouville equation which is used
in statistical physics to describe a gas of particles but here we
use it to describe a single particle.

If the distribution $\rho_0(q,p)$ for $t=0$ is known, we can
consider the Cauchy problem for the equation (\ref{Lio}):
\begin{equation}\label{cau}
\rho |_{t=0}=\rho_0(q,p)\,.
\end{equation}
Let us discuss the case when the initial distribution has the
Gaussian form:
\begin{equation}\label{gauss}
\rho_0 (q,p)=\frac{1}{\pi
ab}e^{-\frac{(q-q_0)^2}{a^2}}e^{-\frac{(p-p_0)^2}{b^2}}\,.
\end{equation}
At sufficiently small values of the parameters $a> 0$ and $b> 0$ the
particle has coordinate and momentum close to the $q_0$ and $p_0$.
For this distribution the average value of the coordinates and
momentum are:
\begin{equation}\label{mean1}
\overline {q}=\int q\rho_0 (q,p)dqdp=q_0\,,~~\overline{p}=\int
p\rho_0 (q,p)dqdp=p_0\,,
\end{equation}
and dispersion
\begin{equation}\label{disp1}
\Delta q^2=\overline{(q-\overline {q})^2}=\frac{1}{2}a^2,~~\Delta
p^2=\overline{(p-\overline {p})^2}=\frac{1}{2}b^2\,.
\end{equation}

\section{Free Motion} Consider first the case of the free motion of
the particle when $V=0$. In this case the equation (\ref{Lio}) has
the form
\begin{equation}\label{free}
\frac{\partial \rho}{\partial t}=-\frac{p}{m}\frac{\partial
\rho}{\partial q}
\end{equation}
and the solution of the Cauchy problem is
\begin{equation}\label{solfree} \rho (q,p,t)=\rho_0
(q-\frac{p}{m}t,p)\,.
\end{equation}
Using expressions (\ref{gauss}), (\ref{solfree}),
\begin{equation}\label{gauss2}
\rho (q,p,t)=\frac{1}{\pi
ab}\exp\{-\frac{(q-q_0-\frac{p}{m}t)^2}{a^2}-\frac{(p-p_0)^2}{b^2}\}\,,
\end{equation}
we get the time dependent distribution of coordinates:
\begin{equation}\label{coor1}
\rho_{c} (q,t)=\int \rho (q,p,t)dp=\frac{1}{\sqrt{\pi}
\sqrt{a^2+\frac{b^2t^2}{m^2}}}\exp\{-\frac{(q-q_0-
\frac{p_0}{m}t)^2}{(a^2+\frac{b^2t^2}{m^2})}\}\,,
\end{equation}
while the distribution of momenta is
\begin{equation}\label{mom1}
\rho_{m} (p,t)=\int \rho
(q,p,t)dq=\frac{1}{\sqrt{\pi}b}e^{-\frac{(p-p_0)^2}{b^2}}.
\end{equation}
Thus, for the free particle the distribution of the particle
momentum with the passage of time does not change, and the
distribution of the coordinates change. There is, as one says in
quantum mechanics, the spreading of the wave packet. From
(\ref{coor1}) it follows that the dispersion $\Delta q^2$ increases
with time:
\begin{equation}\label{disp2}
\Delta q^2(t)=\frac{1}{2}(a^2+\frac{b^2t^2}{m^2})\,.
\end{equation}
Even if the particle was arbitrarily well localized ($a^2$ is
arbitrarily small) at $t = 0$, then at sufficiently large times $t$
 the localization of the particle becomes meaningless, there is a
{\it delocalization} of the particle.

\section{Newton's Equation for the Average Coordinate}

In the functional approach to classical mechanics there is no
ordinary picture of an individual trajectory of a particle. The
starting equation is the dynamic equation (\ref{Lio}) for the
distribution function, rather than the Newton equation.

What role can play the Newton equation in the functional approach?
We show that the average  coordinate for the free particle in the
functional approach satisfies the Newton equation. Indeed, the
average coordinate and momentum for the free particles have the form
\begin{equation}\label{coor2} \overline {q}(t)=\int
q\rho_{c} (q,t)dq=q_0+\frac{p_0}{m}t\,,~~\overline{p}(t)=\int
p\rho_m(p,t)dp=p_0\,.
\end{equation}
Hence we get
\begin{equation}\label{coor3}
\frac{d^2}{dt^2}\overline {q}(t)=0\,,
\end{equation}
i.e. we have Newton's equation for the average coordinates.

We also have Hamilton's equations for the average values of the
coordinate and momentum:
\begin{equation}\label{ham1}
\dot{\overline {q}}=\frac{\partial H}{\partial \overline{p}}\,,~~
\dot{\overline {p}}=-\frac{\partial H}{\partial \overline{q}}\,,
\end{equation}
where the Hamiltonian $H=H(\overline {q},\overline {p})$ for the
free particle has the form
\begin{equation}\label{ham2}
H=\frac{\overline{p}^2}{2m}\,.
\end{equation}
Note that in the functional mechanics the Newton equation for the
average  coordinates is obtained only for the free particle or for
quadratic Hamiltonians with a Gaussian initial distribution
function. For a more general case there are corrections to Newton's
equations, as discussed below.

We discussed the spreading of  Gaussian distribution functions.
Similar results are obtained for the distribution functions of other
forms, if they describe in some sense localized  coordinates and
momenta  at the initial time.

\section{Comparison with Quantum Mechanics}
Compare the evolutions of Gaussian distribution functions in
functional classical mechanics and in quantum mechanics for the
motion of particles along a straight line. The scene of work for the
functional classical mechanics is $ L ^ 2 (\mathbb {R} ^ 2) $ (or $
L ^ 1 (\mathbb {R} ^ 2) $), and for quantum mechanics - $ L ^ 2
(\mathbb {R} ^ 1) $.

The Schrodinger equation for a free quantum particle on a line
reads:
\begin{equation}\label{quan}
i\hbar\frac{\partial \psi}{\partial
t}=-\frac{\hbar^2}{2m}\frac{\partial^2\psi}{\partial x^2}\,.
\end{equation}
Here $\psi=\psi(x,t)$  is the wave function and $\hbar$ is the
Planck constant. The density of the distribution function for the
Gaussian wave function has the form (see, for example \cite{Flu})
\begin{equation}\label{quan2}
\rho_q(x,t)=|\psi (x,t)|^2=\frac{1}{\sqrt{\pi}
\sqrt{a^2+\frac{\hbar^2t^2}{a^2m^2}}}\exp\{-\frac{(x-x_0-
\frac{p_0}{m}t)^2}{(a^2+\frac{\hbar^2t^2}{a^2m^2})}\}\,.
\end{equation}
We find that the distribution functions in functional classical and
in quantum mechanics (\ref {coor1}) and (\ref {quan2}) coincide, if
we set
\begin{equation}\label{unc}
a^2 b^2=\hbar^2\,.
\end{equation}
If the condition (\ref {unc}) is satisfied then the Wigner function
$ W (q, p, t) $ \cite{Sch} for $ \psi $ corresponds to the classical
distribution function (\ref {gauss2})\,, $ W (q, p, t) = \rho (q, p,
t) $ \,.

The problem of spreading of the quantum wave packet in dealing with
the potential barrier is considered in \cite {Din}.

Gaussian wave functions on the line are coherent or compressed
states. The compressed states on the interval are considered in
\cite {VT}.

\section {Liouville Equation and the Newton Equation}

 In the functional classical mechanics  the motion of a particle
 along the stright line is described by the
  Liouville equation (\ref {Lio}). A more general Liouville equation
  on the manifold $ \Gamma $ with coordinates
  $ x = (x^1 ,..., x^k) $ has the form
\begin{equation}\label{Lio2a}
\frac{\partial \rho}{\partial t}+\sum_{i=1}^k
\frac{\partial}{\partial x^i}(\rho v^i)=0\,.
\end{equation}
Here $\rho=\rho (x,t)$ is the density function and
$v=v(x)=(v^1,...,v^k)$ - vector field on $\Gamma$. The solution of
the Cauchy problem for the equation  (\ref{Lio2a}) with initial data
\begin{equation}\label{Lio3}
\rho|_{t=0}=\rho_0(x)
\end{equation}
might be written in the form
\begin{equation}\label{Lio4}
\rho (x,t)=\rho_0(\varphi_{-t}(x))\,.
\end{equation}
Here $\varphi_t(x)$ is a phase flow along the solutions of the
characteristic equation
\begin{equation}\label{char2}
\dot{x}=v(x)\,.
\end{equation}
 In particular, if  $k=2n,$ and $M=M^n$  is a smooth manifold,
the phase space $\Gamma=T^*M$ is a cotangent bundle, $H=H(q,p)$  is
a Hamiltonian function on $\Gamma$, then the Liouville equation has
the form
\begin{equation}\label{LioH2}
\frac{\partial \rho}{\partial t}+\sum_{i=1}^n [\frac{\partial
H}{\partial p^i}\frac{\partial \rho}{\partial q^i}-\frac{\partial
H}{\partial q^i}\frac{\partial \rho}{\partial p^i}]=0\,.
\end{equation}
The Liouville measure $d\mu =dqdp$ is invariant under the phase flow
$\varphi_t$.

Classical dynamical system in the functional approach to mechanics
is a stochastic process $\xi(t)=\xi(t;q,p)=\varphi_{t}(q,p)$ which
takes values in $\Gamma$ and with the probabilistic measure
$dP(q,p)=\rho_0(q,p)dqdp$. Correlation functions have the form
\begin{equation}\label{correll}
<\xi_{i_1}(t_1)...\xi_{i_s}(t_s)>=\int
\xi_{i_1}(t_1;q,p)...\xi_{i_s}(t_s;q,p)\rho_0 (q,p)dqdp\,.
\end{equation}
Here $i_1,...,i_s=1,...,k$.

It is assumed usually that the energy surfaces $\{H=const\}$ are
compact.

A system from $N$  particles in the  3-dimensional space has the
phase space $\mathbb{R}^{6N}$ with coordinates
$q=(\mathbf{q}_1,...,\mathbf{q}_N),~
p=(\mathbf{p}_1,...,\mathbf{p}_N),~\mathbf{q}_i=(q_i^1,q_i^2,q_i^3),~
\mathbf{p}_i=(p_i^1,p_i^2,p_i^3),~i=1,...,N$ and it is described by
the Liouville equation for the function $\rho=\rho (q,p,t)$
\begin{equation}\label{Lio5}
\frac{\partial \rho}{\partial t}=\sum_{i,\alpha}( \frac{\partial
V(q)}{\partial q_i^{\alpha}}\frac{\partial \rho}{\partial
p_i^{\alpha}}-\frac{p_i^{\alpha}}{m_i}\frac{\partial \rho}{\partial
q_i^{\alpha}})\,.
\end{equation}
Here summation goes on $ i = 1 ,..., N, \, \alpha = 1,2,3. $ The
characteristics equations  for (\ref {Lio5}) are  Hamilton's
equations
\begin{equation}\label{char3}
\dot{q_i^{\alpha}}=\frac{\partial H}{\partial
p_i^{\alpha}},\,\,\dot{ p_i^{\alpha}}=-\frac{\partial H}{\partial
q_i^{\alpha}}\,,
\end{equation}
where the Hamiltonian is
\begin{equation}\label{ham3}
H=\sum_{i}\frac{\mathbf{p}_i^2}{2m_i}+V(q)\,.
\end{equation}
Emphasize here again that the Hamilton equations (\ref{char3}) in
the current functional approach to the mechanics do not describe
directly the motion of particles, and they are only the
characteristic equations for the Liouville equation (\ref {Lio5})
which has a physical meaning. The Liouville equation (\ref {Lio5})
can be written as
\begin{equation}\label{Lio6}
\frac{\partial \rho}{\partial t}=\{H,\rho\}\,,
\end{equation}
where the Poisson bracket
\begin{equation}\label{poib}
\{H,\rho\}=\sum_{i,\alpha}( \frac{\partial H}{\partial
q_i^{\alpha}}\frac{\partial \rho}{\partial
p_i^{\alpha}}-\frac{\partial H}{\partial p_i^{\alpha}}\frac{\partial
\rho}{\partial q_i^{\alpha}})\,.
\end{equation}
Criteria for essential self-adjointness of the Liouville operator in
the Hilbert space $L^2(\mathbb{R}^{6N})$ are given in \cite{RS}.

\section{Corrections to Newton's Equations}
In section 4, it was noted that for the free particle in the
functional approach to classical mechanics the averages coordinates
and momenta satisfy the Newton equations. However, when there is a
nonlinear  interaction, then in functional approach corrections  to
the Newton's equations appear.

Consider the motion of a particle along the line in the functional
mechanics. Average value $ \overline {f} $ of the function on the
phase space $ f = f (q, p) $ at time $ t $ is given by the integral
 (\ref{int1})
\begin{equation}\label{defcor}
\overline{f}(t)=<f(t)>=\int f(q,p)\rho (q,p,t)dqdp\,.
\end{equation}
Here  $\rho (q,p,t)$ has the form (\ref{Lio4})
\begin{equation}\label{corr}
\rho (q,p,t)=\rho_0(\varphi_{-t}(q,p))\,.
\end{equation}
By making the replacement of variables,  subject to the invariance
of the Liouville measure, we get
\begin{equation}\label{correcti}
<f(t)>=\int f(q,p)\rho (q,p,t)dqdp=\int f(\varphi_{t}(q,p))\rho_0
(q,p)dqdp\,.
\end{equation}

Let us take
\begin{equation}\label{correc}
\rho_0(q,p)=\delta_{\epsilon}(q-q_0)\delta_{\epsilon}(p-p_0)\,,
\end{equation}
where
\begin{equation}\label{corr1}
\delta_{\epsilon}(q)=\frac{1}{\sqrt{\pi}\epsilon}e^{-q^2/\epsilon^2}\,,
\end{equation}
 $q\in \mathbb{R} , \, \epsilon >0$.

 Let us show that in the limit $\epsilon \to 0$ we obtain
 the Newton (Hamilton) equations:
\begin{equation}\label{correc10} \lim_{\epsilon\to 0}<f(t)>
=f(\varphi_t(q_0,p_0))\,.
\end{equation}

{\bf Proposition 1}. {\it Let the function $f(q,p)$ in the
expression (\ref{defcor})  be continous and integrable, and $\rho_0$
has the form (\ref{correc}). Then}
\begin{equation}\label{correc1}
\lim_{\epsilon\to 0}\int f(q,p)\rho
(q,p,t)dqdp=f(\varphi_t(q_0,p_0))\,.
\end{equation}
{\bf Proof}.
 Functions $\delta_{\epsilon}(q)$ form a
  $\delta$-sequence in $D^{'}(\mathbb{R})$ \cite{Vla}. Hence we obtain
\begin{equation}\label{correc2}
\lim_{\epsilon\to 0}\int f((q,p))\rho (q,p,t)dqdp=\lim_{\epsilon\to
0}\int
f(\varphi_{t}(q,p))\delta_{\epsilon}(q-q_0)\delta_{\epsilon}(p-p_0)=
f(\varphi_t(q_0,p_0))\,,
\end{equation}
that was required to prove.

Now calculate the corrections to the solution of the equation of
Newton. In functional mechanics consider the equation, see
(\ref{Lio})\,,
\begin{equation}\label{Lio2}
\frac{\partial \rho}{\partial t}=-p\frac{\partial \rho}{\partial
q}+\lambda q^2 \frac{\partial \rho}{\partial p}\,.
\end{equation}
Here  $\lambda$ is a small parameter and we set the mass $m=1$. The
characteristic equations have the form of the following Hamilton
(Newton) equations:
\begin{equation}\label{corr3}
\dot{p}(t)+\lambda q (t)^2=0\,,\,\, \dot{q}(t)=p(t)\,.
\end{equation}
Solution of these equations with the initial data
\begin{equation}\label{corr4}
q(0)=q,\,\,\dot{q}(0)=p
\end{equation}
 for small $t$ has the form
\begin{equation}\label{corr5}
(q(t), p(t))=\varphi_t
(q,p)=(q+pt-\frac{\lambda}{2}q^2t^2+...,\,p-\lambda q^2 t+...)
\end{equation}
Use the asymptotic expansion $\delta_{\epsilon}(q)$ in
$D^{'}(\mathbb{R})$ for $\epsilon \to 0$, compare \cite{ALV, PV}:
\begin{equation}\label{corr2}
\delta_{\epsilon}(q)=\delta
(q)+\frac{\epsilon^2}{4}\delta^{''}(q)+...\,,
\end{equation}
then for  $\epsilon \to 0$ we obtain corrections to the Newton
dynamics:
\begin{equation}\label{corr6}
<q(t)>=\int(q+pt-\frac{\lambda}{2}q^2t^2+...)[\delta
(q-q_0)+\frac{\epsilon^2}{4}\delta^{''}(q-q_0)+...]
\end{equation}
$$
\cdot [\delta
(p-p_0)+\frac{\epsilon^2}{4}\delta^{''}(p-p_0)+...]dqdp=q_0+p_0
t-\frac{\lambda}{2}q_0^2t^2 -\frac{\lambda}{4}\epsilon^2t^2\,.
$$
Denoting the Newton solution
$$
q_{\rm Newton}(t)=q_0+p_0 t-\frac{\lambda}{2}q_0^2t^2\,,
$$
we obtain for small $\epsilon, t$ and $\lambda$:
\begin{equation}\label{corr6sm}<q(t)>=
q_{\rm Newton}(t)-\frac{\lambda}{4}\epsilon^2t^2\,.
\end{equation}

Here $-\frac{\lambda}{4}\epsilon^2t^2$  is the correction to the
Newton solution  received within the functional approach to
classical mechanics with the initial Gaussian distribution function.
If we choose a different initial distribution we get correction of
another form.

We have proved

{\bf Proposition 2}.  {\it In the functional approach to mechanics
the first correction at $\epsilon$ to the Newton dynamics for small
$ t$ and $\lambda$ for equation (\ref{corr3}) has the form
(\ref{corr6sm})}.

Note that in the functional approach to mechanics instead of the
usual Newton equation
\begin{equation}\label{New1}
m\frac{d^2}{dt^2}q(t)=F(q)\,,
\end{equation}
where $F(q)$ is a force, we obtain
\begin{equation}\label{Newfunc}
m\frac{d^2}{dt^2}<q(t)>=<F(q)(t)> \,.
\end{equation}
Indeed, multiplying the equation
\begin{equation}\label{Lio2n}
\frac{\partial \rho}{\partial t}=-\frac{p}{m}\frac{\partial
\rho}{\partial q}-F(q) \frac{\partial \rho}{\partial p}\,.
\end{equation}
by $q$ and making integration over $p$ and $q$ and then integrating
by parts, we get
\begin{equation}\label{Lio2n1}
\frac{d}{dt}<q(t)>=\frac{<p(t)>}{m}\,.
\end{equation}
Similarly, multiplying the equation (\ref {Lio2n}) by $ p $ and
integrating on $ p $ and $ q $ and then integrating by parts, we get
\begin{equation}\label{Lio2n1}
\frac{d}{dt}<p(t)>=<F(q)(t)>\,,
\end{equation}
which gives (\ref{Newfunc}).

The task of calculating the corrections at $ \epsilon $ for Newton's
equation for mean values is similar to the problem of calculating
semiclassical corrections in quantum mechanics \cite{MF, FY, VT}.

\section{Time Reversal}

\subsection{Reversibility in classical mechanics}
Let us present a  famous discourse  which proves reversibility of
the dynamics in classical mechanics.  From the symmetry of Newton's
equations upon the replacement the time $ t $ for $-t $ it follows
that if in the system there exists some motion, then it is possible
also the  reverse motion, i.e. such motion, in which the system
passes same states in the phase space in the reverse order. Indeed,
let the function $ x (t) $ satisfies the Newton equation
\begin{equation}\label{newtrev}
\ddot{x}(t)=F(x(t))
\end{equation}
with initial data
\begin{equation}\label{newtreve}
x(0)=x_0,~~\dot{x}(0)=v_0.
\end{equation}
We denote the corresponding solution by $$x(t)=\Phi (t;
x_0,v_0)\,.$$ We fix $T>0$ and let us reverse the motion of the
particle at some moment in time  $T$ by reversing its velocity, i.e.
let us consider the solution $y(t)$ of the Newton equation
\begin{equation}\label{newtrev1} \ddot{y}(t)=F(y(t))
\end{equation}
with the following initial data:
\begin{equation}\label{newtrev2}
y(0)=x(T),~~\dot{y}(0)=-\dot{x}(T)\,.
\end{equation}
Then it is easy to see that at the    time moment  $T$ we get
\begin{equation}\label{newtrev3} y(T)=x_0,~~\dot{y}(T)=-v_0\,,
\end{equation}
i.e. the particle comes back to the initial point with the inverse
velocity. To prove the relation (\ref{newtrev3}) it is enough to
note that the solution of equation(\ref{newtrev1}) with initial data
(\ref{newtrev2}) has the form
$$
y(t)=\Phi (T-t; x_0,v_0)
$$
and use the relations (\ref{newtreve}).

Let us notice that these arguments about reversibility of motion in
the classical mechanics used not only symmetry of the Newton
equation  concerning time reversibility, but also the fact that a
state of the particle in the classical mechanics at some instant of
time is completely  characterized by  two parameters - co-ordinate
$x $ and speed $v $. Reversibility of the motion in classical
mechanics means reversibility of the motion along a given
trajectory.

As it was discussed above, the notion of  an individual trajectory
of a particle has no physical sense. In  reality we deal with a
bunch of trajectories or probability distribution. In the functional
classical mechanics the state of the particle  is characterized not
by the two numerical parameters, but the distribution function $
\rho =\rho (q, p, t) $. In the following subsection it will be
shown, how it leads to delocalization and irreversibility.

\subsection {Irreversibility in the functional mechanics}

The considered reversibility of motion in  classical mechanics deals
with the individual trajectory. In the functional mechanics the
concept of the  individual trajectory of the particle has no direct
physical sense. Instead, the state of the particle  is described by
the distribution function $ \rho =\rho (q, p, t) $ which satisfies
the Liouville equation (\ref{Lio})
\begin{equation}\label{Lioirr}
\frac{\partial \rho}{\partial t}=-\frac{p}{m}\frac{\partial
\rho}{\partial q}+\frac{\partial V(q)}{\partial q} \frac{\partial
\rho}{\partial p}\,.
\end{equation}
The Liouville equation is invariant under the replacement $t $ to
$-t $: if $ \rho =\rho (q, p, t) $  is the solution of the equation
 (\ref {Lioirr}),  then $ \sigma (q, p, t) = \rho (q,-p,-t) $ - also its
solution. However this symmetry does not mean reversibility of the
motion of a particle in the functional approach to mechanics, since
the state of the particle  is described there by the  distribution
function and the phenomenon of delocalization takes place.

In this way we obtain an answer to the  arguments of Loschmidt and
Poincar\'e - Zermelo. Indeed, to reverse the particle motion at the
time moment $t=T $ as it is proposed in the Loschmidt argument, it
is necessary make the co-ordinate and momentum measurement. But it
will change the distribution $ \rho (q, p, T) $. Further, it is
necessary to prepare such condition of the particle that its
evolution back in time would lead to the initial distribution $
\rho_0$ that is difficult since the delocalization takes place. We
will need something even better than  Maxwell's demon.

For free particle the delocalization leads to the increasing of
dispersion $\Delta q^2$ with time (\ref{disp2}):
\begin{equation}
\Delta q^2(t)=\frac{1}{2}(a^2+\frac{b^2t^2}{m^2})\,.
\end{equation}
Notice that the similar phenomena takes place for the Brownian
motion $B(t)$ which has dispersion $t$ \cite{PR, BS}.

Concerning the Zermelo argument related with the Poincar\'e
recurrence theorem we note that this argument can not be applied to
the functional mechanics because this argument is based on the
notion of individual trajectory. In the functional mechanics the
state of the system is characterized by the distribution function
and here the mean values might irreversibly tend to some limits
without contradiction with the Poincar\'e theorem as it will be
shown in the next section.

The Poincar\'e theorem is not applicable to the bunch of
trajectories or even to two trajectories as it follows from the
Lyapunov theory: if two points are situated in some small region of
the phase space then they are not necessary come back to this region
by moving along their trajectories.

\subsection{Mixing and weak limit}
The state $\rho_t=\rho_t (x)$ on the compact phase space $\Gamma$ is
called mixing if its weak limit at  $t\to\infty$ is a constant,
$$
\lim_{t\to\infty}\rho_t (x)=const\,.
$$
More precisely, a dynamical system $(\Gamma, \varphi_t,d\mu)$ has
the mixing property \cite{Sin,KH} if
\begin{equation}\label{mix}
\lim_{t\to\infty}<f,U_t g>=\int\bar{f}d\mu\cdot\int g d\mu
\end{equation}
for every $f,g\in L^2(\Gamma)$. Here $U_t g(x)=g(\varphi_t(x))$. For
the mixing systems the bunch of trajectories is spreading over the
phase space, hence in the functional mechanics we have
irreversibility.

The method of the weak limit which generalizes the Poincar\'e
results and which can be applied to a wide class of dynamical
systems is developed in  \cite{Koz1,Koz2}.

Connection with the irreversibility problem can be explained on the
following example. Let us consider the function of two real
variables
$$ F(t,p)=e^{itp}f(p)\,,
$$
where $f(p)$ is an integrable function. It is clear that the
function $F(t,p)$ is periodic in $t$ if  $p$ is fixed and it has no
limit as $t\to\infty$. However, if we integrate the function
$F(t,p)$ over $p$,
$$
F(t)=\int e^{itp}f(p)dp\,,
$$
then we get the function $F(t)$ which already has the limit (by the
Riemann - Lebesgue lemma):
$$
\lim_{t\to\infty}F(t)=0\,.
$$

\section{Dynamics of a Particle in a Box}

Dynamics of collisionless continuous medium in a box with reflecting
walls is considered in \cite{Poi3, Koz1, Koz2}. This studied
asymptotics of solutions of Liouville equation. In functional
approach to mechanics, we interpret the solution of the Liouville
equation as described the dynamics of a single particle. Here we
consider this model in the classical and  also in the quantum
version for the special case of Gaussian initial data.

\subsection {Dynamics of a classical particle in a box}
Consider the motion of a free particle on the interval with the
reflective ends. Using the method of reflections \cite{Vla}, the
solution of the Liouville equation(\ref{free})
$$
 \frac{\partial \rho}{\partial
t}=-\frac{p}{m}\frac{\partial \rho}{\partial q}
$$
on the interval $0\leq q\leq 1$ with the reflective ends we write as
\begin{equation}\label{refl}
\rho (q,p,t)=\sum_{n=-\infty}^{\infty} [\rho_0
(q-\frac{p}{m}t+2n,p)+\rho_0 (-q+\frac{p}{m}t+2n,-p)]\,,
\end{equation}
where it is assumed that the function $\rho_0$ has the Gaussian form
(\ref{gauss}).

One can show that for the distribution for coordinates
\begin{equation}\label{distr}
\rho_c (q,t)=\int \rho (q,p,t)dp
\end{equation}
one gets the uniform limiting distribution (pointwise limit):
$$
\lim_{t\to\infty}\rho_c (q,t)=1\,.
$$
For the distribution of the absolute values of momenta ($p>0$)
$$
\rho_a (p,t)=\rho_m (p,t)+\rho_m (-p,t)\,,
$$
where
$$
\rho_m (p,t)=\int_{0}^{1} \rho (q,p,t)dq\,,
$$
as $t\to\infty$ we get the distribution of the Maxwell type (but not
the Maxwell distribution):
$$
\lim_{t\to\infty}\rho_a
(p,t)=\frac{1}{\sqrt{\pi}b}[e^{-\frac{(p-p_0)^2}{b^2}}
+e^{-\frac{(p+p_0)^2}{b^2}}]\,.
$$
\subsection{Dynamics of a quantum particle in a box}

 The Schrodinger equation for free quantum particle on the interval
  $0\leq x\leq
1$ with reflecting ends has the form
\begin{equation}\label{quansegm}
i\hbar\frac{\partial \phi}{\partial
t}=-\frac{\hbar^2}{2m}\frac{\partial^2\phi}{\partial x^2}
\end{equation}
with the boundary conditions
$$
\phi (0,t)=0,~~\phi (1,t)=0,~~t\in\mathbb{R}\,.
$$
Solution of this boundary problem can be written as follows:
$$
\phi (x,t)=\sum_{n=-\infty}^{\infty}[\psi(x+2n,t)-\psi(-x+2n,t)]\,,
$$
where $\psi (x,t)$ is some solution of the Schrodinger equation. If
we choose the function $\psi (x,t)$ in the form, corresponding to
the distribution (\ref{quan2}), then one can show that in the
semiclassical limit for the probability density $|\phi(x,t)|^2$ the
leading term is the classical distribution $\rho_c(x,t)$
(\ref{distr}).

\section{Conclusions}
In this paper the functional formulation of classical mechanics is
suggested which is based not on the notion of an individual
trajectory of the particle but on the distribution function on the
phase space.

The fundamental equation of the microscopic dynamics in the proposed
functional approach is not the Newton equation but the Liouville
equation for the distribution function of a single particle.
Solutions of the Liouville equation have the property of
delocalization which accounts for irreversibility. It is shown that
the Newton equation in this approach appears as an approximate
equation describing the dynamics of the average values of the
positions and momenta for not too long time intervals. Corrections
to the Newton equation are computed.

Interesting problems related with applications of the functional
formulation of mechanics to  statistical mechanics, to singularities
in cosmology and black holes, and new interpretation of quantum
mechanics we hope to consider in further works.

\section{Acknowledgements}
The author expresses his sincere thanks to G.A. Alexeev, I.Ya.
Aref`eva, O.V. Groshev, B. Dragovich, E.A. Dynin, M.G. Ivanov, A.Yu.
Khrennikov, V.V. Kozlov, Yu.I. Manin, E.V. Piskovsky, A.S.
Trushechkin, V.A. Zagrebnov, E.I. Zelenov, and participants of the
special seminar on the irreversibility problem NOC MIAN for fruitful
discussions of the fundamental problems of mechanics. The work is
partially supported by grants NS-3224.2008.1, RFBR 08-01-00727-a,
AVTSP 3341 and the program OMN RAS.


\begin{thebibliography}{99}

\bibitem{Bol1} Boltzmann L. Lectures on Gas Theory,
Berkeley, CA: U. of California Press, 1964.

\bibitem{Poi1} Poincar\'e A. ``Le mecanisme et l'experience",
Revue de Metaphysique et de Morale 1, pp. 534-7 (1893); English
translation, Stephen Brush, Kinetic Theory, vol.2, p.203.
\bibitem{Bog} Bogolyubov N.N., Problems of Dynamic Theory
in Statistical Physics, [in Russian], OGIZ (1946).
\bibitem{Kri}  Krylov N.S., Works on the Foundations
of Statistical Physics [in Russian], Akad. Nauk. SSSR, Moscow --
Leningrad (1950).
\bibitem{LL} Landau L.D., Lifshitz E.M.  Statistical physics,
part 1 (3 ed., Pergamon, 1980)
\bibitem{Che} Chester G.V., The theory of irreversible processes.
Reports on Progress in Physics, Volume 26, Number 1, 1963 , pp.
411-472.
\bibitem{Kac} Kac M.  Probability and
Related Topics in Physical Sciences. John Wiley, New York, (1959).
\bibitem{Zub} Zubarev D. N.  Nonequilibrium Statistical Thermodynamics,
  New York, Consultants Bureau, (1974).


\bibitem{Ohya} Ohya M. Some Aspects of Quantum Information Theory
and Their Applications to Irreversible Processes. Rep. Math. Phys.,
vol. 27, pp. 19-47, 1989.

\bibitem{Pri} Prigogine I. Les Lois du Chaos. Flammarion. Paris. 1994.
\bibitem{Koz1} Kozlov V.V. Temperature equilibrium on Gibbs and Poincar\'e,
Moscow-Ijevsk (in Russian), 2002.
\bibitem{Koz2} Kozlov V.V. Gibbs ensembles and nonequilibrium
statistical mechanics, Moscow-Ijevsk (in Russian), 2008.

\bibitem{ALV}
Accardi L., Lu Y. G. and Volovich I. Quantum Theory and Its
Stochastic Limit. Berlin: Springer, 2002.


\bibitem{Gin} Ginzburg V.L. Which problems in physics
are most important and interesting in the beginning of XXI century?
In: V.L.Ginzburg. On science, on myself and others, (in Russian),
Fizmatlit, 2003, pp. 11 - 74.

\bibitem{Gal} Gallavotti G. Fluctuation relation,
fluctuation theorem, thermostats and
entropy creation in non equilibrium statistical physics.
 arXiv:cond-mat/0612061.

\bibitem{Fey} Feynman R. The character of physical law,
A series of lectures recorded by the ÂÂÑ at Cornell University USA,
Cox and Wyman Ltd., London, 1965.


\bibitem{Leb} Lebowitz J.L. From Time-symmetric Microscopic
Dynamics to Time-asymmetric Macroscopic Behavior: An Overview.
arXiv:0709.0724.

\bibitem{Gol} Goldstein S. Boltzmann's approach to statistical
mechanics, in: J. Bricmont, et al. (Eds). Chance in Physics:
Foundations and Perspectives, Lecture Notes in Physics, 574,
Springer-Verlag, Berlin. 2001. P. 39-68.

\bibitem{Bri} Bricmont J. Science of Chaos or Chaos in Science?
in: The Flight from Science and Reason, Annals of the N.Y. Academy
of Sciences 775, p. 131-182. 1996.

\bibitem{Kol}  Kolmogorov A.N. The general theory of
dynamical systems and classical mechanics.
Proceedings of the International Congress
of Mathematicians (Amsterdam, 1954),
Vol. 1, pages 315-333, North Holland,
 Amsterdam, 1957 [in Russian].
 English translation as Appendix D
 in R.H. Abraham, Foundations of Mechanics,
 pages 263-279. Benjamin, 1967.
 Reprinted as Appendix in R.H.
 Abraham and J.E. Marsden, Foundations of Mechanics,
Second Edition, pages 741-757. Benjamin / Cummings 1978.

\bibitem{Arn} Arnold V.I. Mathematical methods of classical mechanics,
Springer-Verlag, 1978.

\bibitem{DNF} Dubrovin B.A., Fomenko A.T., Novikov S.P.. Modern Geometry.
Methods and Applications. Springer-Verlag, GTM 93, Part 1, 1984.


\bibitem{Ano}  Anosov D.V. Geodesic flows on closed
Riemann manifolds with negative curvature, Proc. Steklov Inst. Math.
Vol. 90, 1960.
\bibitem{Sin}  Sinai Ya.G. Introduction to
ergodic theory. Princeton Univ. Press, Princeton, New Jersey, 1977.

\bibitem{Bol5} Boltzmann L. Ober die Beziehung
eines allgemeine mechanischen Satzes zum zweiten Hauptsatze der
Warmetheorie, Sitzungsberichte Akad. Wiss., Vienna, part II, 75,
67-73 (1877); English transl: Stephen Brush, Kinetic Theory, vol. 2,
p.188.

\bibitem{Fri} Friedman A.  "Uber die Krummung des Raumes".
Zeitschrift fur Physik,
10 (1): 377–386, (1922).


\bibitem{Lin} Linde A.D. Inflation and Quantum Cosmology.
Academic Press, Boston 1990.

\bibitem{NV}  Nieuwenhuizen Th.M. ,  Volovich I.V.
Role of Various Entropies in the Black Hole Information Loss
Problem. in "Beyond the Quantum", eds. Th.M. Nieuwenhuizen, V.
Spicka, B. Mehmani, M. J. Aghdami, and A. Yu. Khrennikov (World
Scientific, 2007); arXiv:hep-th/0507272.


\bibitem{Poi3} Poincare H. , "Remarks on the Kinetic Theory of Gases,"
in Selected Works (Nauka, Moscow, 1974), Vol. 3.


\bibitem{Poi2} Poincare H. (1904), "L'etat actuel et
l'avenir de la physique mathematique", Bulletin des sciences
mathematiques 28 (2): 302–324, 1904.  English translation in
Poincare, Henri (1904), "The present and the future of mathematical
physics", Bull. Amer. Math. Soc. (2000) 37: 25–38.



\bibitem{KT}  Kozlov V.V.,  Treschev D.V., "Fine-grained
and coarse-grained entropy in problems of statistical mechanics",
TMF, 151:1 (2007), 120–137.

\bibitem{KH} Katok A. and Hasselblatt B.
Introduction to the Modern Theory of Dynamical Systems. New York:
Cambridge University Press.

\bibitem{Vol1}
Volovich I. V.  "Number theory as the ultimate physical theory", \\
Preprint No. TH 4781/87, CERN, Geneva, 1987.


\bibitem{Vol2}
 Volovich I.V. $p$-adic string. //
Class. Quant. Grav. V. 4.  P. L83--L87. 1987.

\bibitem{Man} Manin Yu.I. Reflections on arithmetical physics,
In: Manin Yuri I. Mathematics as Metaphor: Selected Essays of Yuri
I. Manin. Providence, R.I.: American Mathematical Society, 2007, pp.
149--155.

\bibitem{VVZ}
 Vladimirov V.S.,   Volovich  I.V.,   Zelenov E.I. $p$--Adic Analysis
and Mathematical Physics. Singapore: World Scientific, 1994.

\bibitem{Khr1}
Khrennikov A. Yu.  Non-Archimedean analysis: quantum paradoxes,
dynamical systems and biological models, Kluwer Acad. Publishers,
Dordreht/Boston/London, 1997.

\bibitem{DKKV}
Dragovich B., Khrennikov  A. Yu.,  Kozyrev S. V. and Volovich I. V.
On p -adic mathematical physics.// P-Adic Numbers, Ultrametric
Analysis, and Applications. Vol. 1. N. 1. P. 1-17. 2009.

\bibitem{Var} Varadarajan V. S.
Multipliers for the symmetry groups of p -adic spacetime.// P-Adic
Numbers, Ultrametric Analysis, and Applications. Vol. 1. N. 1. P.
69-78. 2009.



\bibitem{Zel} Zelenov E. I. Quantum approximation theorem.//
P-Adic Numbers, Ultrametric Analysis, and Applications. Vol. 1. N.
1. P. 88-90. 2009.

\bibitem{Vol3} Volovich I.V.  Quantum Cryptography in Space and Bells Theorem,
in: Foundations of Probability and Physics, Ed. A. Khrennikov, World
Sci.,2001, pp.364-372.

\bibitem{Vol5} Volovich I.V. Seven Principles of Quantum Mechanics,
arXiv: quant-ph/0212126.

\bibitem{Mac} Mackey G.W.  Mathematical foundations of quantum mechanics.
W. A. Benjamin, New York, 1963.



\bibitem{FY}   Faddeev L.D., Yakubovsky O.A. "Lectures on quantum mechanics
for mathematical students", 2-nd edition, Moscow, 2001.

\bibitem{Kli} Klimontovich Yu.L. Statistical physics. Moscow, Nauka, 1982.


\bibitem{LL2}  Landau L.D., Lifshiz E.M. Fluid mechanics.
Pergamon press, 1959.


\bibitem{Flu}  Flugge S. Practical quantum mechanics.
Springer, 1994.

\bibitem{Sch} Schleich, W. P. Quantum optics in phase space.
Wiley-VCH Verlag, 2001.

\bibitem{Din} Dynin E.A. Statistical moments
in quantum tunneling. Soviet physics. Semiconductors. 24:44,
480-481, 1990.

\bibitem{VT} Volovich I.V., Trushechkin A.S.
On quantum compressed states on interval and uncertainty relation
for nanoscopic systems. Proceedings of Steklov Mathematical
Institute, v. 265. pp. 1-31. 2009.

\bibitem{RS} Reed M., Simon B. Methods of Modern Mathematical Physics.
vol.2, Academic press, 1975.

\bibitem{Vla}  Vladimirov V.S.
Equations of Mathematical Physics, Marcel Dekker, New York, 1971.


\bibitem{PV}  Pechen A.N.,  Volovich I.V.
Quantum Multipole Noise and Generalized Quantum Stochastic
Equations. Infinite Dimensional Analysis, Quantum Probability and
Related Topics, Vol. 5, No 4 (2002) 441-464.



\bibitem{MF}
Maslov V.P., Fedoryuk M.V. Semi-classical approximation in Quantum
Mechanics. Dordrecht–Boston–London: D. Reidel Publ., 1981.


\bibitem{PR}  Prokhorov Yu.V. and
 Rosanov Yu.A. Probability Theory. Springer-Verlag, Berlin, 1969.

\bibitem{BS}  Bulinsky A.V. and
 Shiryaev A.N. Theory of Random Processes. Fizmatlit, Moscow,
2003.

\bibitem{Khr2} Khrennikov A.Yu.  Interpretations of probability, VSP Int.
Publ., Utrecht, 1999.
\end{thebibliography}
 \end{document}